\documentclass[table, dvipsnames]{aastex63}
\usepackage{amsmath}
\usepackage{graphicx}
\usepackage[utf8]{inputenc}

\usepackage{enumerate}
\usepackage{verbatim}
\usepackage{todonotes}

\usepackage{hyperref}
\usepackage[capitalize, noabbrev]{cleveref}

\pdfoutput=1

\crefname{section}{\S}{\S\S}

\makeatletter
\usepackage{etoolbox}
\patchcmd\H@refstepcounter{\protected@edef}{\protected@xdef}{}{}
\makeatother

\newcommand{\myemail}{bjoshi5@jhu.edu}

\newcommand{\rst}{\emph{Roman }}
\newcommand{\sn}{SN~Ia }
\newcommand{\sne}{SNe }
\newcommand{\totalsn}{1698}
\newcommand{\degree}{$^{\circ}$}
\definecolor{lgray}{gray}{0.8}

\received{May 2022}
\accepted{Nov 2022}

\begin{document}

\title{High-Precision Redshifts for Type Ia Supernovae with the \emph{Nancy Grace Roman Space Telescope} P127 Prism}
\shorttitle{Roman Prism Simulations of SNe Ia}
\shortauthors{Joshi et al.}

\author[0000-0002-7593-8584]{Bhavin A.\ Joshi}
\altaffiliation{Corresponding Author}
\email{\myemail}
\affiliation{Department of Physics and Astronomy, Johns Hopkins University, Baltimore, MD 21218, USA}
\author[0000-0002-7756-4440]{Louis-Gregory Strolger}
\affiliation{Department of Physics and Astronomy, Johns Hopkins University, Baltimore, MD 21218, USA}
\affiliation{Space Telescope Science Institute, 3700 San Martin Drive, Baltimore, MD 21218, USA}
\author{Russell E.\ Ryan, Jr.}
\affiliation{Space Telescope Science Institute, 3700 San Martin Drive, Baltimore, MD 21218, USA}
\author[0000-0003-3460-0103]{Alexei V. Filippenko}
\affiliation{Department of Astronomy, University of California, Berkeley, CA 94720-3411, USA}
\author[0000-0002-0476-4206]{Rebekah Hounsell}
\affiliation{University of Maryland, Baltimore County, Baltimore, MD 21250, USA}
\affiliation{NASA Goddard Space Flight Center, Greenbelt, MD 20771, USA}
\author{Patrick L. Kelly}
\affiliation{School of Physics and Astronomy, University of Minnesota, Minneapolis, MN 55455, USA}
\author{Richard Kessler}
\affiliation{Kavli Institute for Cosmological Physics, University of Chicago, Chicago, IL 60637, USA}
\author{Phillip Macias}
\affiliation{Department of Astronomy and Astrophysics, University of California, Santa Cruz, CA 95064, USA}
\author[0000-0002-1873-8973]{Benjamin Rose}
\affiliation{Department of Physics, Duke University, 120 Science Drive, Durham, NC, 27708, USA}
\author{Daniel Scolnic}
\affiliation{Department of Physics, Duke University, 120 Science Drive, Durham, NC, 27708, USA}

\begin{abstract}
We present results from simulating slitless spectroscopic observations with the {\it Nancy Grace Roman Space Telescope's} ({\it Roman}) Wide-Field Instrument (WFI) P127 prism spanning 0.75\,$\mu m$ to 1.8\,$\mu m$. We quantify the efficiency of recovered Type Ia supernovae (SNe~Ia) redshifts, as a function of P127 prism exposure time, to guide planning for future observing programs with the \emph{Roman} prism. Generating the two-dimensional dispersed images and extracting one-dimensional spectra is done with the slitless spectroscopy package \texttt{pyLINEAR} along with custom-written software. From the analysis of \totalsn\ simulated SN~Ia P127 prism spectra, we show the efficiency of recovering SN redshifts to $z\lesssim3.0$, highlighting the exceptional sensitivity of the \rst P127 prism. Redshift recovery is assessed by setting a requirement of $\sigma_z = (\left|z - z_\mathrm{true} \right|)/(1+z) \leq 0.01$. We find that 3\,hr exposures are sufficient for meeting this requirement, for $\gtrsim 50\%$ of the sample of mock SNe~Ia at $z\approx2$ and within $\pm5$ days of rest-frame maximum light in the optical. We also show that a 1\,hr integration of \rst can achieve the same precision in completeness to a depth of $24.4 \pm 0.06$ AB mag (or $z\lesssim 1$). Implications for cosmological studies with \rst P127 prism spectra of SNe~Ia are also discussed.
\end{abstract}

\section{Introduction}
\label{sec:intro}

The {\it Nancy Grace Roman Space Telescope} ({\it Roman}) will, through its core community High Latitude Time Domain Survey (HLTDS), observe on the order of $10^4$ Type Ia supernovae \citep[SNe~Ia; e.g.,][]{Hounsell:2018vu, Akeson2019}. Most of these SNe will be within redshifts $0.5 < z < 2$, and are required to be observed with sufficient depth per epoch to enable accurate measurements of dark energy properties that significantly improve the dark energy figure of merit \citep[FoM;][]{Spergel:2015wr,Hounsell:2018vu}. To accomplish these goals, the two Supernova Science Investigation Teams (SITs) have provided a jointly recommended reference survey in \citet{Rose:2021wk}.

An important assumption in \citet{Rose:2021wk} is that the redshifts for \textit{all} SNe~Ia (or host galaxies) are well measured, i.e., on the order of a tenth of a percent, and it is assumed that \rst spectroscopic elements will be used as a part of the {\it Roman} HTLDS strategy to obtain those redshifts. The \rst Wide-Field Instrument (WFI), with a field-of-view of 0.281\,deg$^2$, is expected to have two slitless-spectroscopy spectral elements: a low-resolution prism (P127; $R\approx70$--170; $0.75 \lesssim \lambda\,[\mu m] \lesssim 1.8$) and a slightly higher-resolution grism (G150; $R\approx435$--865; $1.0 \lesssim \lambda\,[\mu m] \lesssim 1.93$).\footnote{ \url{https://roman.gsfc.nasa.gov/science/WFI_technical.html} provides more complete details on the current \rst WFI design.} Compared to the grism, the \rst prism, while lower resolution, is $\sim 3$\,mag more sensitive than the \rst grism for a 1\,hr exposure making it better suited for analysis of fainter SNe with broad absorption features (\cref{tab:hst_roman_compare}).

As part of the SIT effort, the primary objective of this work is to evaluate the utility of the prism spectra, particularly the influence of the exposure time on the signal-to-noise ratio (S/N) and redshift accuracy. These results are needed to optimize the spectroscopic component of the HLTDS and future observations with the \rst prism. Current estimates for the HLTDS design involve some (as yet undecided) combination of prism-to-imaging time, along with coaddition of spectra taken from multiple visits, as well as time series analysis \citep{Saunders2018}. Based on estimates of current \rst WFI/P127 capabilities, we provide here a base platform from which future considerations of prism exposures for the HLTDS can continue, while also giving a broader context for other programs that involve the \rst prism. We analyze the redshift recovery for a single long-exposure visit (which can still include multiple position angles) where all data for a given \sn are taken such that spectral or luminosity variations can be neglected (over the timescale of the visit).

With simulated broadband filter images of a single \rst HLTDS pointing (hereafter referred to as direct images) in the WFI/F106 filter as input \citep{Wang2022}, our approach involves simulating two-dimensional (2D) spectroscopic images (dispersed images) through the \rst prism, and analyzing the extracted one-dimensional (1D) spectra, as will be done for real \rst data. This paper focuses primarily on the recovery of \sn redshifts, although our fitting pipeline also attempts to constrain \sn phase (relative to peak) and dust extinction along the line of sight. The simulation products and the associated codes will be publicly available at the IPAC-hosted simulations page\footnote{\url{https://roman.ipac.caltech.edu/sims/Simulations_csv.html}} and GitHub\footnote{\url{https://github.com/bajoshi/roman-slitless}}, respectively. 
Future papers will focus on recovery of (a) other \sn properties (e.g., \ion{Si}{2} velocities) while also including coadded shorter exposures from multiple visits at different phases such that the SN brightness and spectral evolution are measured, (b) deblending of \sn and host galaxy spectra, and (c) (host-)galaxy stellar-population parameters from their prism spectra.

This work is structured as follows: \cref{sec:simmethods} goes into details of the simulation and analysis methods, \cref{sec:simmodels} discusses the models used for SNe~Ia, galaxies, and stars; \cref{sec:blending} describes how host galaxy light and SN light are blended for SNe that overlap with their host galaxy in our simulation; \cref{sec:2dsim} and \cref{sec:x1d} provide an overview of the \texttt{pyLINEAR} package used to simulate the 2D dispersed images and extract the 1D spectra; and \cref{sec:fitting} describes the fitting process applied to the extracted 1D spectra. Finally, \cref{sec:efficiency} discusses the results obtained from our fitting pipeline, particularly with regards to the efficiency of recovering redshifts. The cosmological model assumed in this paper is flat with $\Omega_\mathrm{M} = 1 - \Omega_\Lambda = 0.3$ and H$_0 = 70$\,km\,s$^{-1}$\,Mpc$^{-1}$, and all magnitudes are in the AB system \citep{Oke1983}.

\section{Simulation and Analysis Methods}
\label{sec:simmethods}
\begin{figure}
    \centering
    \includegraphics[width=\textwidth]{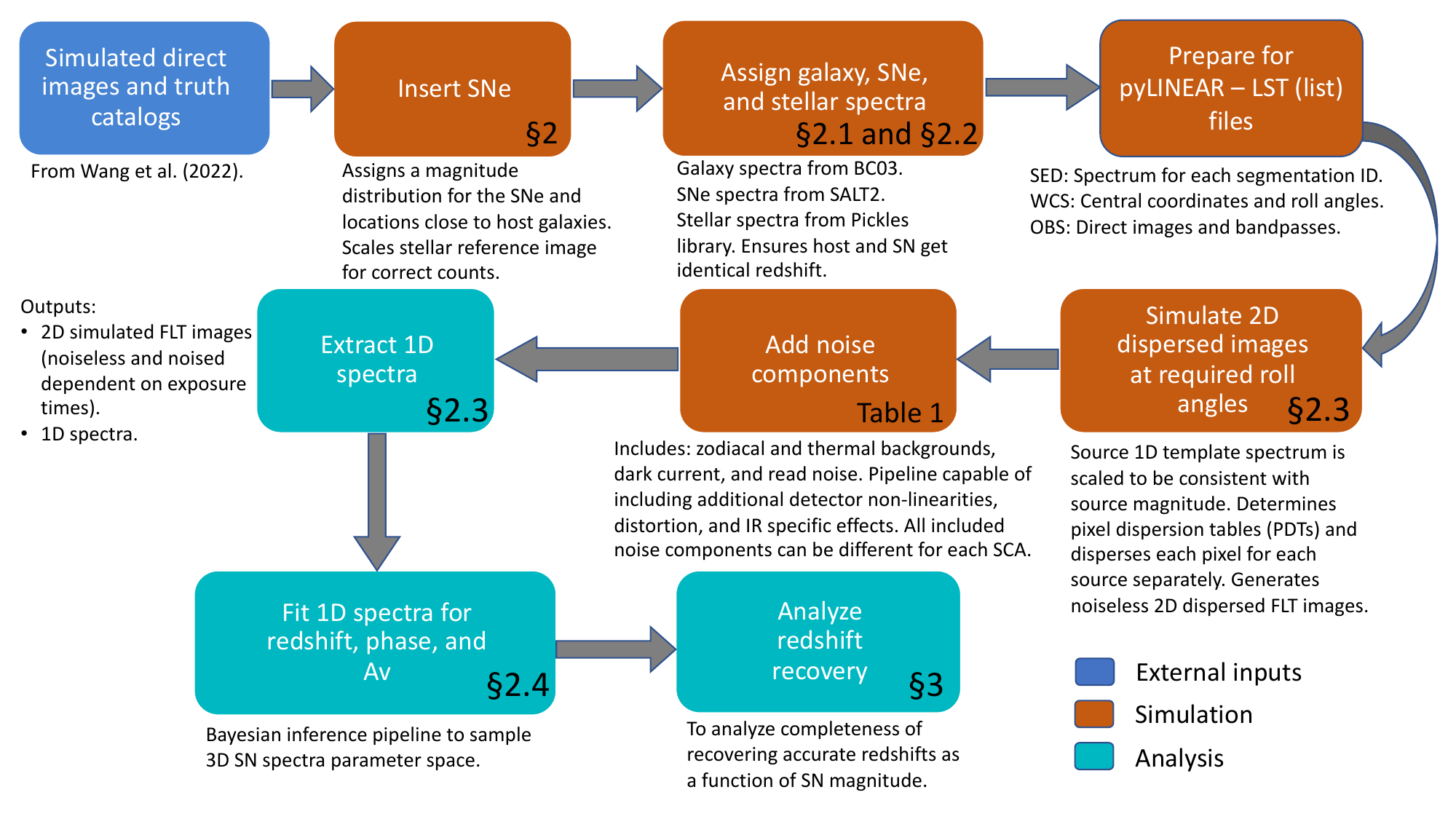}
    \caption{Schematic representation of the steps in our simulation pipeline. A short description of the details in each step is also given. Each box also provides, if applicable, a section/table number where it is described in detail in the text.}
    \label{fig:flowchart}
\end{figure}

Redshifts and distance moduli are the two independently-measured quantities for SNe~Ia that allow astronomers to constrain cosmological models. Our goal here is to analyze the accuracy of redshifts recovered from simulated 1D P127 spectra of SNe~Ia, while keeping the simulation as realistic as possible. A schematic of the simulation pipeline, as well as the inputs and products, is shown in \cref{fig:flowchart}. 

Our process consists of the following steps which includes an end-to-end simulation (steps i and ii) and analysis of the simulation products (steps iii and iv): 
\begin{enumerate}[(i)]
    \item insert \sne as point sources into simulated direct images and assign \sne and (host-)galaxies spectra (\cref{sec:simmodels} and \cref{sec:blending}), 
    \item generate dispersed images at required roll angles and exposure times for the \rst WFI/P127 prism, (\cref{sec:2dsim})
    \item extract 1D slitless spectra of all detected sources (\cref{sec:x1d}), and
    \item recovery of SNe~Ia and galaxy stellar population parameters through a Markov Chain Monte Carlo (MCMC; \cref{sec:fitting}) fitting of the ``observed'' slitless spectroscopy.
\end{enumerate}

The simulation is based on one pointing, of all 18 WFI detectors, generated at three roll angles with total exposure times of 1200, 3600, and 10,800\,s. The roll angles we use are 0\degree, 70\degree, and 140\degree, measured east of north. The exposure times are built up by adding repeated observations assuming that the longest individual exposure is limited to 900 seconds.
Given that detector-specific details are not yet known for {\it Roman,} our code runs the simulation on a single detector and repeats the simulation 18 times with a different scene on the sky each time (where ``scene'' is defined by the shapes, sizes, locations, and brightness for all objects within the field-of-view).
We also limit the number of randomly-inserted \sne into each detector to a range of $90 \leq N \leq 100$ (i.e., $<5$\% of total sources which are typically $\sim 2000$), so as to not artificially introduce excessive spectral overlap. In total, our simulation contains \totalsn\ SNe~Ia.

\begin{figure}
    \centering
    \includegraphics[width=0.85\textwidth]{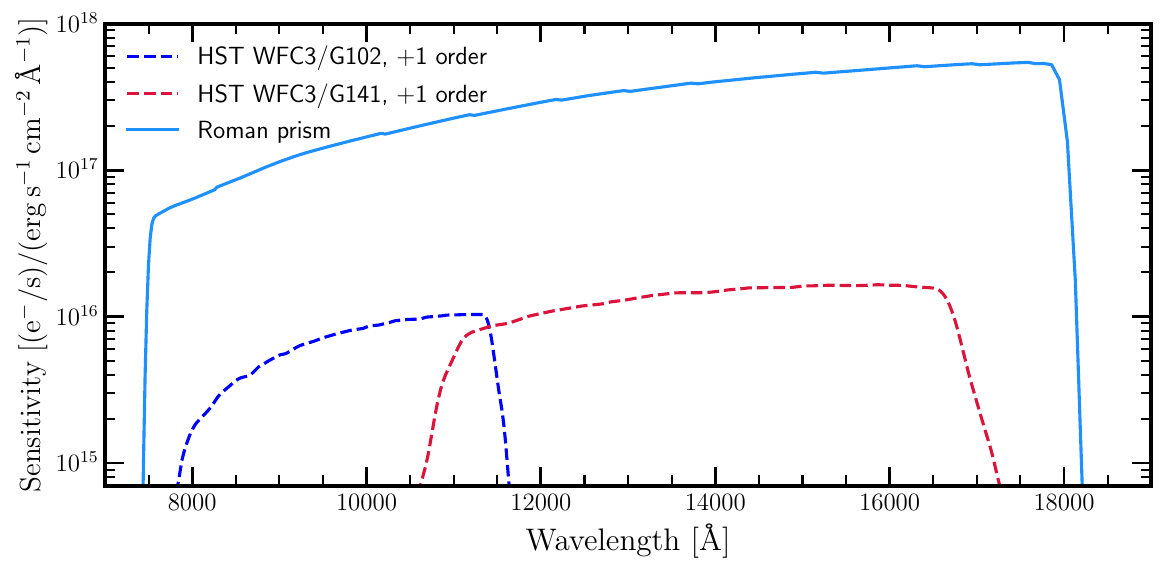}
    \caption{Estimate of the sensitivity of {\it Roman} P127 prism, along with the {\it HST} Wide Field Camera 3 IR grisms G102 and G141, in the +1 order.}
    \label{fig:sensitivity}
\end{figure}

\begin{deluxetable*}{c c c >{\columncolor{lgray}}c c}
\tablecaption{Comparison of {\it HST} and \rst slitless elements
\label{tab:hst_roman_compare}}
\tablehead{
Optics/Detector characteristic & \multicolumn{2}{c}{HST WFC3/IR grisms} & \multicolumn{2}{c}{Roman WFI} \\
  & G102 & G141 & P127 & G150
}
\startdata
Dark current [$e^-$/pix/s] & 0.05 & 0.05 & 0.0015 & 0.0015\\
Read noise [$e^-$ rms] & $\sim 12$ & $\sim 12$ & $\sim 8$ & $\sim 8$\\
Zodiacal background [$e^-$/pix/s] \tablenotemark{a} & 0.48 & 1.04 & 1.047 & 1.047 \\
Thermal background [$e^-$/pix/s] & 0.02 & 0.02 & 0.064 & 0.064 \\
1\,hr limiting mag\tablenotemark{b} & 22.6 & 22.9 & 23.5 & 20.5\\
Resolution [$\Delta \lambda/\lambda$] & $\sim 210$ & $\sim 130$ & 70--170 & 435--865\\
Wavelength coverage [nm] & 800--1150 & 1075--1700 & 750--1800 & 1000--1930 \\
\enddata
\tablenotetext{a}{The zodiacal light estimates for the {\it HST} grisms vary primarily with Sun angle and time of year \citep{Pirzkal2020}. We therefore use median values for the distribution of zodiacal light levels for the {\it HST} grisms. Assumes a 1.1$\times$ factor for zodiacal light for \emph{Roman} (private communication, Jeffrey Kruk) which is expected to depend on Galactic latitude.}
\tablenotetext{b}{For the {\it HST} grisms, limiting magnitude is defined as 5$\sigma$ per continuum pixel in the $J$ and $H$ bands for G102 and G141, respectively \citep{Kuntschner2010}. For the \rst prism and grism, limiting magnitude is defined as $10\sigma$ per 2\,pixel resolution element in the continuum, at 1.2\,$\mu$m, for a point source.}
\tablecomments{A comparison of relevant properties of the {\it HST} WFC3 IR grisms and the \rst prism and grism, some of which factor into the increased sensitivity of the \rst prism. See \url{https://roman.gsfc.nasa.gov/science/WFI_technical.html} and \url{https://wfirst.ipac.caltech.edu/sims/Param_db.html}, for details on WFI detector performance. See \url{https://www.stsci.edu/hst/instrumentation/wfc3/documentation/grism-resources} for details on the {\it HST} WFC3 grisms. Since the grisms have multiple orders, relevant quantities are quoted here for the +1 order.}
\end{deluxetable*}

Simulating the photoelectron count rate for a given source observed through the \rst prism requires use of the sensitivity curve (\cref{fig:sensitivity}) along with (i) the counts for each pixel associated with the source in the direct image, (ii) a description of the spectral trace for the \rst prism, (iii) a spectrum associated with each source pixel, and (iv) the noise characteristics of the instrumentation (see noise budget components in \cref{tab:hst_roman_compare}). The spectral trace is defined as the location of the centroid of the spectrum along the spatial direction \citep[see e.g.,][]{Kummel2009, Pirzkal2016}.
\cref{tab:hst_roman_compare} shows a comparison of properties between the {\it Hubble Space Telescope (HST)} Wide Field Camera 3 (WFC3) infrared (IR) grisms and the \rst prism and grism. \cref{tab:hst_roman_compare} also shows the noise terms included in our simulation. However, it does not include IR detector-specific effects, such as flux-dependent sensitivity and interpixel capacitance, which we expect to have a smaller effect on the prism spectra compared to the terms already included. This table also shows that the primary reason for preferring the \rst prism over the grism for observing SNe~Ia is the $\sim 3$\,mag deeper sensitivity of the prism, which is particularly helpful for estimating redshifts to fainter SNe~Ia given the broad absorption features typical of \sn spectra. Another significant reason for the increased sensitivity of the \rst prism is the lack of dispersion orders. Since the grisms disperse light into multiple orders and the prism disperses light into a single order, the grisms lose a significant fraction of the available light; typically only the +1 order spectrum from a grism is analyzed given that gratings are blazed to maximize throughput to the +1 order.

\cref{fig:sensitivity} shows a comparison between the sensitivities of the \rst prism and {\it HST} WFC3 IR grisms G102 and G141. According to current best estimates of sensitivity, the \rst prism is a factor of $\gtrsim 20$ more sensitive than {\it HST}/WFC3 IR grisms at all wavelengths covered by the {\it HST} grisms. The prism's performance is expected to result in $10\sigma$ per pixel in a 1\,hr exposure of a point source with $m_{F106}$ = 23.5\,mag AB\footnote{\url{https://roman.gsfc.nasa.gov/instruments_and_capabilities.html}} (although we recover $\sim 6\sigma$ per pixel from analyzing our simulations).

\subsection{Stellar Population and SALT2 Models}
\label{sec:simmodels}
Prior to dispersing the simulated direct images we assign each source in the direct images a model spectrum. Briefly, the simulated direct images contain stars and galaxies and while they also contain SNe~Ia, the rates employed result in too few SNe for our analysis (for more details we refer the reader to \citealt{Wang2022} and references therein).

Therefore, we begin by overlaying additional SNe into the direct images as point sources. 
Typical SNe~Ia simulations begin by selecting a random redshift from a rate model, and then computing \sn model magnitudes at each phase. To simulate an appropriate magnitude range, we simulate with a different approach in which random magnitudes are chosen in the range $22.4 \leq m_{F106}\, [\mathrm{AB\ mag}] \leq 28.7$, and the corresponding redshift is determined for each magnitude (corresponding to SNe~Ia approximately within $0.5 \lesssim z \lesssim 3.0$). We overlay the SNe into the direct images with a synthetic apparent magnitude chosen first instead of choosing a redshift first and then letting the apparent magnitude be decided by the cosmology, \sn phase, and line of sight dust attenuation. This approach is motivated by the need to acquire a large enough sample of fainter SNe, because we expect the accuracy of inferred redshifts to deteriorate for fainter SNe, while still completing the simulations within a realistic time-frame, owing to the computationally expensive nature of the simulations. Choosing the apparent magnitude first allows us to have a well-known magnitude distribution for the overlaid SNe~Ia. For the \sn model template, we employ an extension of the SALT2 \sn model that extends into the near-IR \citep{Guy2007, Pierel2018}. This model provides rest-frame SN~Ia spectra from 1700\,\AA\ to 2.5\,$\mu$m, and with phases $-19$\,days to $+50$\,days relative to optical peak. The SALT2 model employed here is a fiducial model, i.e., it assumes $x1 = c = 0$. While this model does not fully capture the diversity in SNe~Ia, it still serves our main purpose of the assessing the redshift recovery for a typical SN Ia at any given redshift. This work is largely focused on analyzing how the S/N affects redshift recovery, for which  our model for an average SN Ia at each redshift appears sufficient (see \cref{sec:fitting} and the discussion in \cref{sec:efficiency}). The diversity of SNe Ia and the analysis of other SN Ia subtypes, which are known to have variable intrinsic brightness and different spectral features (e.g., 91bg or 91T), will be addressed in a future work.

The probability distribution function (PDF) for the magnitude distribution follows a power law ($P(x, a) = a x^{a-1}$ with $a=1.2$), ensuring a distribution of SN magnitudes that allows for a slightly larger number of fainter SNe for analysis. To obtain a statistically significant sample of SNe~Ia for analysis we insert a relatively large number of \sne into each detector (as previously mentioned $\sim 95$ \sne per detector). While this rate model is not consistent with the number density of SNe that will be observed in a single visit, our decision is again driven by the need to obtain a large number of SN prism spectra for analysis.
Each SN is assigned a spectrum with a randomly chosen phase within $\pm 5$\,days from peak.  
Next, we apply dust extinction $A_V$ following the \citet{Calzetti2000} law, chosen from $0.0 \leq A_V \leq 3.0$\,mag, with an exponentially declining PDF weighted in favor of lower $A_V$ values ($P(x, \lambda) = \lambda\,e^{-\lambda x}$ with $\lambda=2$), in order to simulate realistic dust extinction from local environments of SNe, following \citet{Jha1999} and \citet{Holwerda2015}.
Finally, we apply cosmological redshift to the inserted SN spectrum by redshifting the wavelength grid and scaling the spectrum flux so that the synthetic magnitude matches the already assigned SN magnitude.

For (host-) galaxies we use the \citet{Bruzual2003} GALAXEV library of stellar-population models. Each galaxy is assigned a composite stellar population with a delayed-$\tau$ star-formation-rate (SFR) model (SFR $\propto\, t\,\mathrm{exp}(-t/\tau)$) for the star-formation history, along with a stellar mass of $9.0 \leq \mathrm{log( M_s \,[M_\odot])} \leq 12.0$, a redshift of $0 < z \leq 3.0$ (if it is an SN host galaxy, then the SN and host are assigned the same redshift), and finally a dust extinction with $A_V$ randomly chosen from $0.0 \leq A_V \leq 5.0$\,mag.
For Galactic stellar point sources in the images we assign spectra from the Pickles stellar library, that provide stellar spectra from 1150\,\AA\ to 25,000\,\AA\ \citep{Pickles1998}. The Pickles library provides 131 flux-calibrated stellar templates for stellar types within the range O5--M2, with the stellar type being chosen randomly in our simulations.

\subsection{Blending of host galaxy and Supernova Light}
\label{sec:blending}
A significant advantage of slitless spectroscopy is the ability to obtain a spectrum for all objects within the field-of-view (FoV).
This also introduces significant data-reduction challenges, particularly with disentangling spectra from objects that are very close to each other and/or lie along the dispersion direction such that their spectra overlap. In traditional slitless spectroscopy parlance, the term contamination usually refers to the unwanted overlap of one or more spectra on the spectrum of the object of interest. Generally this effect can be mitigated by taking data at multiple roll angles so that there is at least one roll angle within which the spectrum of interest contains little to no contamination. 

\begin{figure*}[t!]
    \centering
    \includegraphics[width=0.9\textwidth]{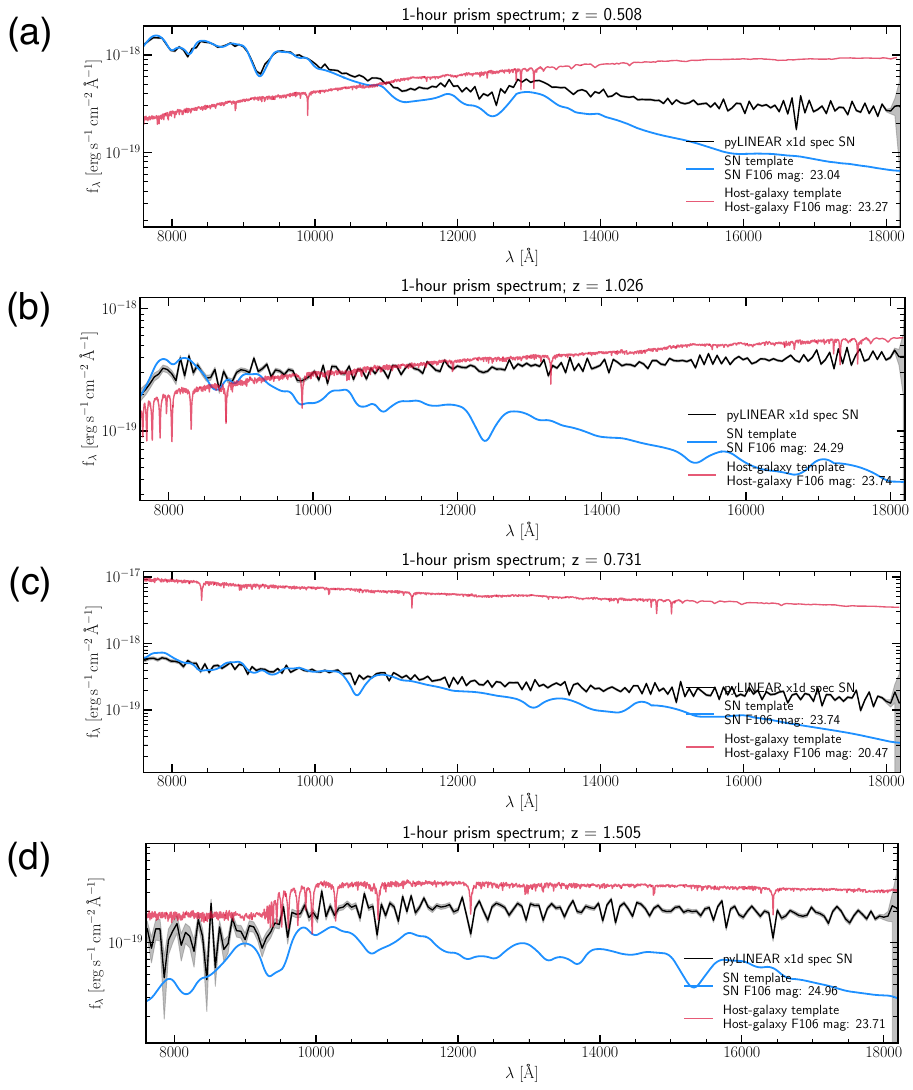}
    \caption{Examples of SNe~Ia prism spectra that are blended with host galaxy spectra. The input high-resolution template spectra are in blue for the SN and red for the host galaxy. Extracted SN spectra are in black with gray errors. Host galaxy and SN redshifts, as well as F106 magnitudes, are shown in the legend. From top to bottom the extent of blending increases. Panel (a) shows an example of mild blending --- SN~Ia features are easily identifiable and host galaxy features are largely absent in the extracted SN spectrum. Panels (b) and (c) show moderate blending --- both SN~Ia and host galaxy features can be identified in the SN spectrum. Panel (d) shows extreme blending between SN and host galaxy --- SN~Ia features are largely absent and host galaxy features dominate in the SN spectrum. This ``classification'' of the extent of blending (as mild, moderate, or extreme) is qualitative --- we do not attempt here to quantify the extent of the blending.
    }
\label{fig:host_blending}
\end{figure*}

In the specific case of slitless spectroscopic observations of an SN~Ia and its host galaxy, an additional concern is the blending of light from the host galaxy and the SN --- that is, within pixels that overlap between the SN and its host galaxy, the total flux (and therefore the spectrum from that pixel) is a combination of host and SN. This introduces a minimum floor of contamination (referred to as blending below to distinguish from the traditional notion of contamination) that will be present in all observations of the SN~Ia regardless of the roll angle. This situation is analogous to that of taking slitless spectra of a galaxy containing one or more star-forming knots --- in this case, the underlying galaxy spectrum is blended with the emission-line spectrum of the star-forming region.

In order to represent this effect within our simulations, for each SN source that has an overlap between the host galaxy and SN we carry out the following steps.\\
(1) For each overlapping pixel $(x,y)$, the effective spectrum from the pixel is given by
\begin{equation}
\mathcal{O} = a_{\rm H}\,f^{\rm H}_\lambda + a_{\rm SN}\,f^{\rm SN}_\lambda,
\end{equation}
where $a_{\rm H}$ and $a_{\rm SN}$ denote linear combination coefficients, and $f^{\rm H}_\lambda$ and $f^{\rm SN}_\lambda$ are the spectra of the host galaxy and SN, respectively. The coefficients $a_{\rm H}$ and $a_{\rm SN}$ are fractions of the total counts within a pixel that belong to the host galaxy and SN, respectively. We also make the simplifying assumption (for all sources) that the input spectrum remains constant over the spatial extent of the source.\\
(2) The final spectrum of the SN containing overlap with the host is given by a weighted sum of the spectra from overlapping pixels (i.e., blended SN and host spectra) and non-overlapping pixels (i.e., pure SN~Ia spectra):
\begin{equation}
\label{eqn:blend}
\mathcal{S} = \Sigma_i a_i \mathcal{O}_i + \Sigma_j b_j \mathcal{N}_j\,,
\end{equation}
where $\mathcal{S}$ is the final blended spectrum of the SN and ($\mathcal{N}$)$\mathcal{O}$ denote (non-)overlapping pixels. The indices, i and j, sum over SN pixels that overlap and do not overlap with the host galaxy, respectively. The weights $a_i$ and $b_i$ are computed to ensure that the relative flux contribution from each pixel toward the ``effective'' source count, which includes both host galaxy and SN counts, is reflected in the final spectrum. This final blended spectrum becomes the effective 1D spectral template for the SN, i.e., each pixel associated with the SN, in the segmentation map, is assigned the spectrum $\mathcal{S}$. The effective 1D spectral template is then passed to \texttt{pyLINEAR} for simulating 2D dispersed images as described in \cref{sec:2dsim}.

\cref{fig:host_blending} shows examples of blended SN~Ia and host galaxy spectra. Extracted spectra (the extraction process is described in \cref{sec:x1d}) are effectively the weighted sum of the SN and host input templates (\cref{eqn:blend}). The extent of blending depends on the number of pixels that overlap between the SN and host galaxy, the exposure time, and their brightness relative to each other. Without any attempt to deblend the host galaxy spectrum from the SN spectrum, our automated fitting pipeline (described in \cref{sec:fitting}) fails to recover the SN redshift in each of the cases shown here. 
Overall, $\sim$54\% of the SNe~Ia in our simulation are affected, to varying degrees, by blended host galaxy light. We arrive at this fraction by counting any SN in our simulation that has one or more pixels that overlap with the host galaxy. This ``blended'' fraction is affected by the placement of the SNe which in our case is at the vertices of the ``bounding box'' of the segmentation pixels of the host galaxy. This effectively simulates placing the SNe~Ia at the effective light radius of the host galaxy. This placement choice avoids completely overwhelming the SN light by the host galaxy light while still allowing a significant fraction of SNe to contain blended light for later analysis.

In practice, for dark energy studies, SNe~Ia that are deeply embedded in host galaxy light would be rejected (if the host light cannot be reliably subtracted) so that they do not affect the inference of cosmological parameters. The total number of SNe~Ia  expected within the \rst cosmological sample has already estimated by various other studies ($\sim10^4$; e.g., \citet{Hounsell:2018vu, Akeson2019}). With regards to host galaxy blended light this sample size will depend on the strictness of any selection cuts for blended host light and the success of any algorithms used to deblend host galaxy light which are being developed for \emph{Roman}. We therefore do not expect that inferred cosmological parameters will significantly depend on the amount of blended host galaxy light, however their statistical errors which depend on the sample size will be affected.

\begin{figure}
    \centering
    \includegraphics[width=0.85\textwidth]{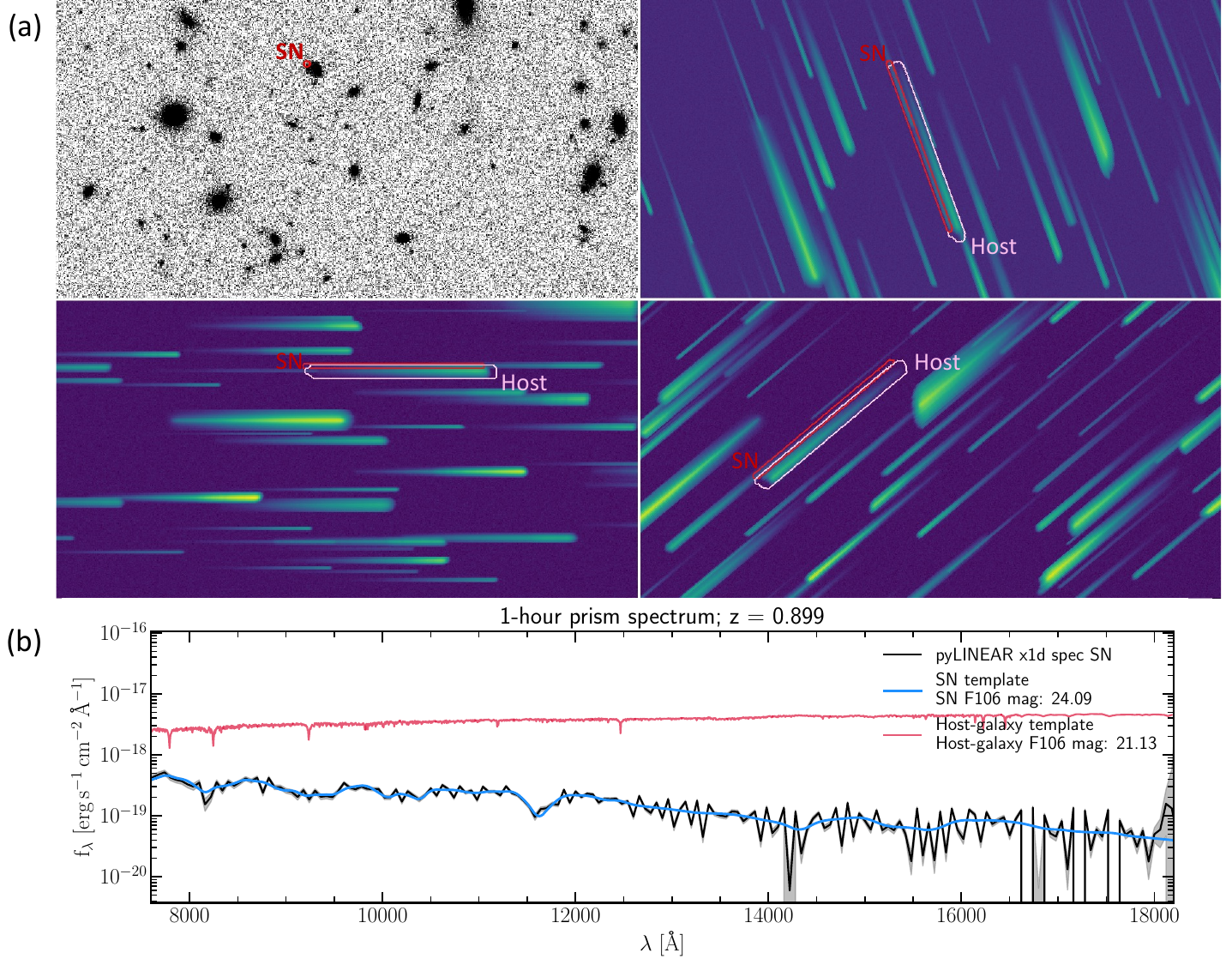}
    \caption{Dispersed prism image cutouts and extracted spectra for an example pair of SN and its host galaxy. The set of panels in (a) show a portion of the direct image (for a single detector, $\approx$ 600 $\times$ 450 pixels) and corresponding dispersed images at our three position angles. The dispersed images also highlight the regions used by \texttt{pyLINEAR} to extract the SN and host galaxy spectra. Panel (b) shows the extracted SN spectrum along with the input templates for the SN and host galaxy. The exposure time for the 2D dispersed images and the 1D extracted spectra shown is 1200 seconds. For this particular example the pixels between the SN and host galaxy do not overlap within the segmentation map.}
    \label{fig:2d_viz}
\end{figure}

\subsection{Simulation of 2D dispersed images}
\label{sec:2dsim}
Here we provide a brief overview of the process to simulate 2D dispersed images carried out by \texttt{pyLINEAR}. For a detailed explanation of the \texttt{pyLINEAR} algorithm, the reader is referred to \citet{Ryan2018, pylinear-isr}. 

The driving motivations for the algorithm behind \texttt{pyLINEAR} are to employ dispersed data from multiple roll angles to optimally handle spectral contamination due to overlapping spectra, and also to extract spectra with the optimal spectral resolution without sacrificing the S/N. \texttt{pyLINEAR} simulates 2D dispersed images by dispersing 1D high-resolution model spectral templates for each pixel associated with each source in the direct image. It requires a description of the scene on the sky and a description of the spectral trace and sensitivity of the dispersing element being considered.

We modify \texttt{pyLINEAR} defaults to allow the simulation of prism exposures in which the dispersion is strongly dependent on wavelength. Briefly, traditional slitless spectroscopy routines (e.g., aXe, \citealt{Kummel2009}; \texttt{pyLINEAR}) describe the spectral trace with a polynomial that is parameterized by the path length along the trace. For describing the spectral trace of the {\it Roman} prism, we build on the foundation of the generalized 2D polynomials introduced by \citet{Pirzkal2017}. These polynomials are a function of the $x$ and $y$ coordinates on the dispersed image and another parameter $t$ (which spans $0 \leq t \leq 1$ and can generally, although not always, be identified with wavelength coverage). One of the quantities required to be computed in simulating 2D spectra is the wavelength $\lambda = f_\lambda (x,y,t)$. For most grisms, this is a first-order equation in $t$. To allow for a variable dispersion with wavelength relation, however, we had to change this to a second-order equation in $t$. This change is nontrivial because the polynomial relation must also be inverted to solve for the coefficients of $t$ that give the known dispersion vs.\ wavelength relation. For readers interested in simulating \rst prism spectra, the configuration file for aXe and \texttt{pyLINEAR} is available from the corresponding author.
\cref{fig:2d_viz} shows an example of simulated 2D dispersed images through the \rst prism and the extracted SN spectrum from a SN and its host galaxy.

\section{Analysis of simulation products}
\label{sec:analysis}

\subsection{Extraction of 1D spectra}
\label{sec:x1d}
The extraction of 1D spectra, which involves solving for the flux at each wavelength for each source, is an overconstrained problem that has more data values than unknown parameters --- that is, the set containing observed information (direct and dispersed images) is far larger than the set of information to be solved for (fluxed spectra for all objects).
\texttt{pyLINEAR} attempts to solve for the set of spectra that simultaneously satisfy all the available dispersed data (i.e., all roll angles and dithers). \texttt{pyLINEAR} defines the system of linear equations that connect the observed data and the unknown fluxes and recasts to a known linear-algebra problem with sparse matrices, $\mathbf{Ax=B}$. Roughly, elements in the matrices $\mathbf{A}$ and $\mathbf{B}$ represent the observed counts and the weighting coefficients that transform them into the fluxed spectra, and $\mathbf{x}$ represents the unknown vector of fluxed spectra for all objects.
This amounts to finding a minimum in the equation,
\begin{equation}
    \chi^2 = \Sigma||\mathbf{Ax - B}||^2.
    \label{eqn:pylinear_axb}
\end{equation}
However, if the matrix $\mathbf{B}$ has noise (which real data does), then the solution to equation \ref{eqn:pylinear_axb} is numerically unstable. Therefore, the actual equation to be solved is modified to the following regularized least squares equation,
\begin{equation}
    \chi^2_\mathrm{true} = \Sigma||\mathbf{Ax - B}||^2 + \mathcal{L}\,\Sigma||\mathbf{x}||^2,
\end{equation}
where, $\mathcal{L}$ is the regularization parameter (see \citealt{Ryan2018} for more details on the regularization).
\texttt{pyLINEAR} employs the LSQR \citep{Paige1982} algorithm for solving the above equation with large sparse matrices.

In contrast to aXe \citep{Kummel2009}, which provides a separate spectrum per object for each roll angle, while leaving the combining of spectra at different roll angles to the user, \texttt{pyLINEAR} provides a single extracted 1D spectrum per object given its algorithmic philosophy of solving for the optimal flux vector that satisfies all the available data. Effectively, the combining of the spectra at different roll angles is done internally within \texttt{pyLINEAR}. This approach naturally handles the problem of overlapping spectra without leaving the contamination subtraction as a separate analysis task. Additionally, \texttt{pyLINEAR} has the advantage of achieving better spectral resolution without sacrificing the S/N in the extracted spectrum.

\begin{figure*}[t!]
\centering
\includegraphics[width=0.98\textwidth]{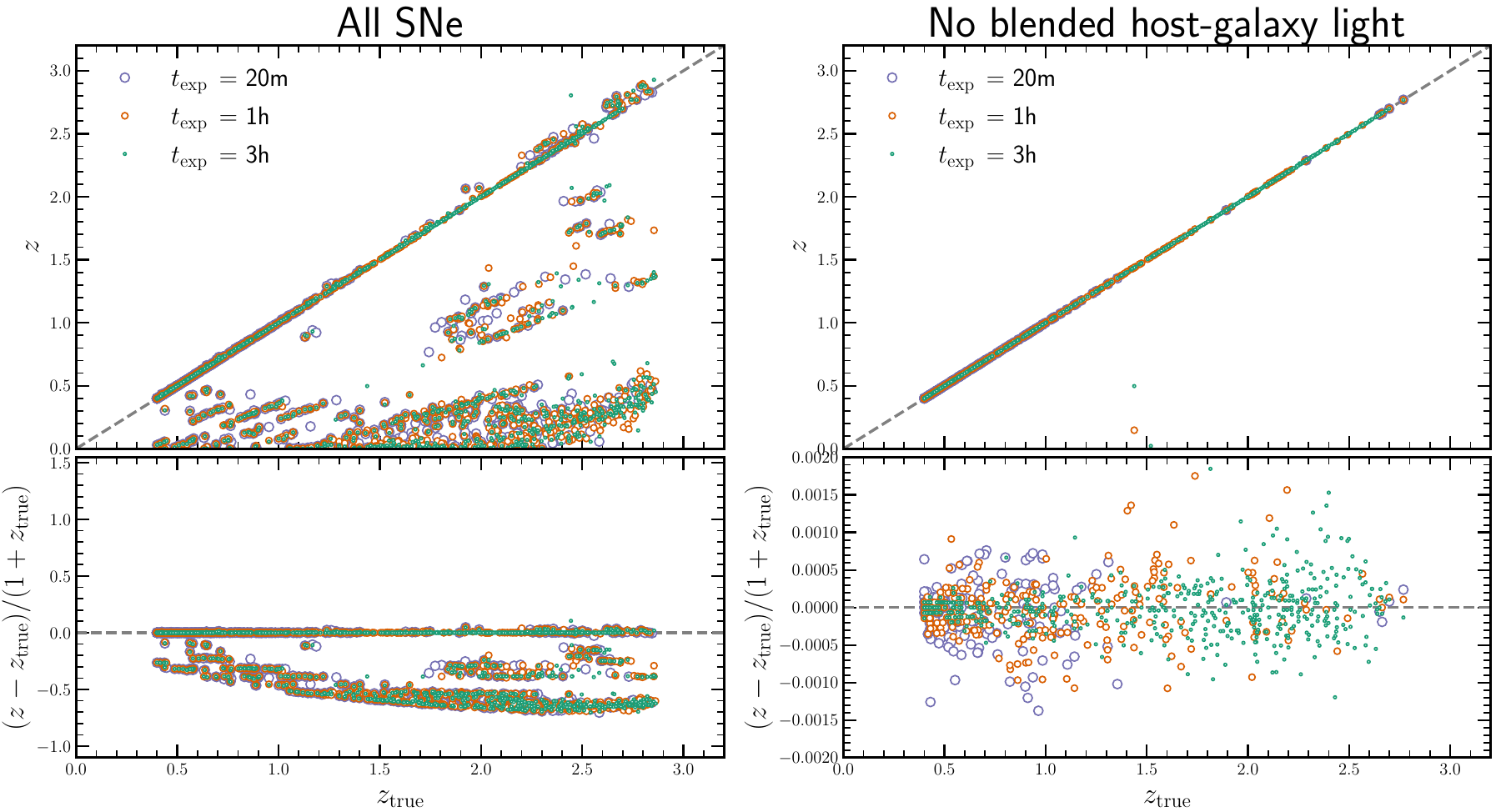}
\caption{The recovery of redshift from our fitting pipeline. Left column shows the recovery results for all the SNe~Ia in our simulation, whereas the right column is restricted to only those SNe whose spectra contain no blended light from the host galaxy. The error bars have been suppressed for clarity. The top panels show the inferred redshift vs. the true redshift and the bottom panels show the residuals defined by $(z_\mathrm{inferred} - z_\mathrm{true})/(1+z_\mathrm{true})$. The colors of the points denote the different exposure times as shown in the legend.}
\label{fig:recovery}
\end{figure*}

\subsection{Parameter Recovery through Bayesian Inference}
\label{sec:fitting}
Extracted 1D spectra are fit for redshift, phase, and dust extinction. It is important to note that we do not fit SALT2 light curve parameters here, $x1$ and $c$. We only fit for the parameters which are typically inferred from spectroscopic data, i.e., redshift, whereas fitting for light curve parameters, at present, requires photometric data.\footnote{In the future, it should be more viable to fit for $x1$ and $c$ from spectral features \citep[e.g., ][]{Mandel2014, Zhao2021}.} Cosmological inference should not be affected by the lack of fitting for $x1$ and $c$, given that redshift is an independent measure needed for the Hubble diagram.
The fitting is done with the Python package {\tt emcee} \citep{Foreman-Mackey2013, Foreman-Mackey2019}, which samples the 3-dimensional (3D) posterior space for \sne models. Our automated fitting pipeline uses 500 walkers and 1000 steps. The walkers use the default ``stretch-move'' within {\tt emcee} for sampling the posterior. We start the walkers at the optimal position indicated by the approximate global minimum.
The minima for the walkers' starting location is found by doing a brute force $\chi^2$ evaluation for each model on a coarse grid of redshift ($\Delta z = 0.01$ within [0.01, 3.0]), phase ($\Delta \mathrm{Phase} = 1$ within [$-19$, 50]), and dust extinction ($\Delta A_V = 0.5$\,mag within [0.0, 3.0]\,mag). 
We employ flat priors on SN phase and dust extinction ($A_V$) with the following ranges: $-19 \leq \mathrm{SN\ Phase\ [Day\ rel.\ to\ peak]} \leq 50$ and $0.0 \leq A_V \leq 5.0$. For the SN redshift we employ a broad Gaussian prior with a mean of 1.2 and a standard deviation of 0.7. The broad Gaussian redshift prior allows down-weighting of very low-$z$ and high-$z$ \sne (while not rejecting them outright). Finally, we require S/N$>$3 for the extracted 1D spectrum prior to fitting. This S/N is computed for the entire prism spectrum using the DER-SNR algorithm of \citet{Stoehr2008}.

\cref{fig:recovery} shows the recovery of \sn redshifts for all SNe in the simulation (left column) and for the subset without host galaxy light (right column). The recovery plots for phases and dust attenuation are shown in the appendix (\cref{fig:recovery_appendix}).
As expected, longer exposure times provide the most accurate inferred quantities. Most of the failed estimates of redshift are caused by host galaxy light blending with SN light. 
As shown in \cref{fig:recovery}, the inferred redshifts are systematically underestimated for SNe whose spectra contain light from the host galaxy, given that these failures are absent in the plot that ignores SNe which contain blended host galaxy light. The failures generally come from three modes. First, the continuum slope of the host galaxy spectrum, when blended with the typically fainter SN spectrum, causes a mismatch between SNe~Ia absorption features. This mismatch is driven by the continuum slope of the host galaxy at the red end which dominates the $\chi^2$ in the SED fitting. The necessity for the fitting pipeline to match the host galaxy continuum slope (instead of the SN continuum slope) pushes the pipeline estimate to underestimated redshifts (and also younger phases, \cref{fig:recovery_appendix}). Second, strong host galaxy absorption features (e.g., 4000\,\AA/Balmer breaks) are misidentified by the fitting pipeline with SN absorption features. This is shown by the clear linear feature below the 1:1 line. Lastly, the cluster of failures at the bottom of the figure are cases where host galaxy light dominates the extracted 1D spectrum.
However, it is worth recalling that the redshift estimates in \cref{fig:recovery} come from fitting only the SNe spectra, and that a host galaxy spectrum can always be observed after the SN light has faded allowing for a clean subtraction (and also allowing for a redshift estimate directly from the host galaxy spectrum in some cases).

In this work, we do not attempt to deblend host galaxy and SN light. Our fitting pipeline proceeds on the assumption that there is no host galaxy light blended into the \sn spectrum, and it fits the \sn spectrum without any attempt to subtract the host galaxy spectrum. We leave this deblending to a future work which will likely involve marginalizing over nuisance parameters that describe the host galaxy spectrum.

\begin{figure*}
    \centering
    \includegraphics[width=0.98\textwidth]{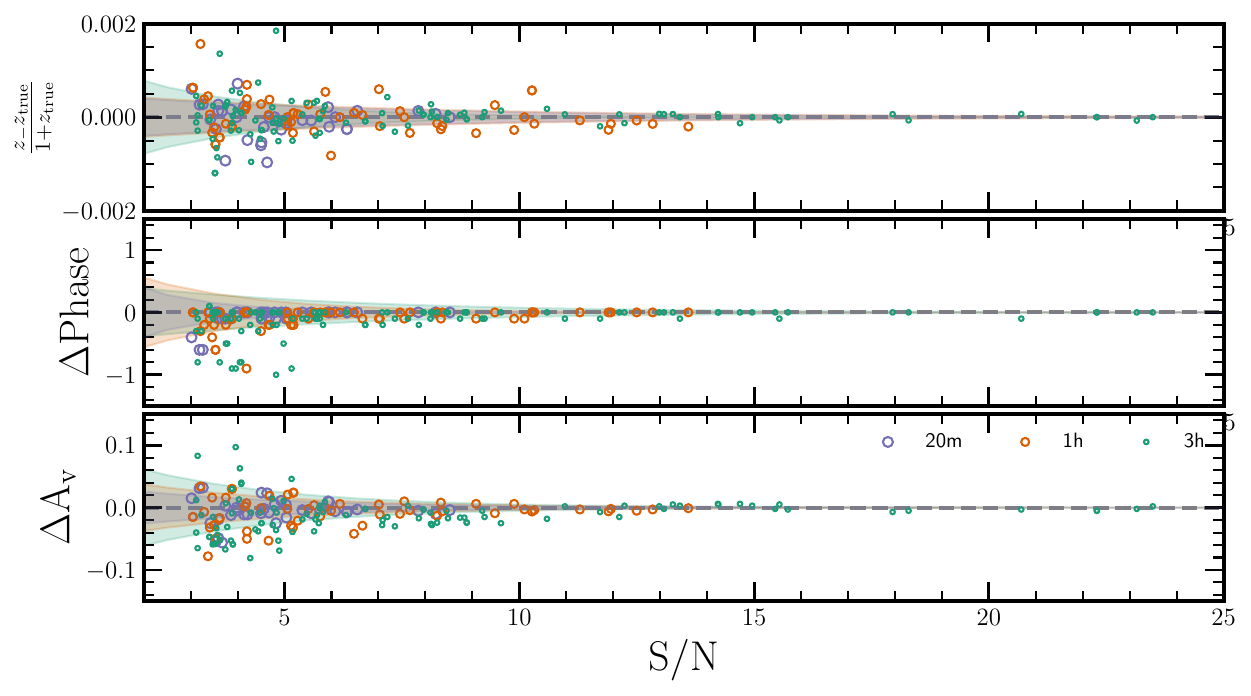}
    \caption{Fitted parameter residuals vs. S/N of observed spectrum for redshift (top), phase (middle), and AV (bottom). Shaded exponential curves enclose $\pm$1$\sigma$ of the residuals. For clarity, only a small subset of the SNe unaffected by blending are shown. The S/N is computed for the entire prism spectrum using the DER-SNR algorithm of \citet{Stoehr2008}. The colors are the same as those used in \cref{fig:recovery}.}
    \label{fig:snr}
\end{figure*}

\section{Redshift Efficiency Dependence on Exposure Time}
\label{sec:efficiency}
In this work, we assess the utility of \rst prism spectra primarily with regard to estimating redshifts. The metric most commonly employed in the literature is to consider statistics of the redshift residuals defined by $\sigma_z = (z_\mathrm{inferred} - z_\mathrm{true})/(1+z_\mathrm{true})$. 

\cref{fig:snr} shows the dependence of the residuals of the recovered parameters on the S/N of the simulated spectrum for different exposure times. For clarity, this plot only shows those SNe from our simulation which do not contain any blending with host galaxy light. For redshift residuals (top row) from low to high S/N, three somewhat distinct S/N regimes can be observed --- S/N $< 5$, $5 \leq$ S/N $\leq 15$, and S/N $> 15$ --- which correspond to increasing accuracy of recovered parameters. Interestingly, while the redshift residuals and the dust-extinction residuals do not exhibit any appreciable bias in their measurements relative to S/N, the recovered phase shows a clear bias toward younger phases (at S/N $\lesssim 5$). We do not yet know the reason for this systematic phase bias. However, we note that a degeneracy between phase and redshift residuals has been observed by other studies \citep[e.g., figure 14 in][]{Blondin2007}. In this context, the recovered phases are underestimated for overestimated redshifts. The bias towards younger recovered phases at lower S/N is perhaps indicative of this degeneracy. While a larger range of simulated phases would allow for a full exploration of parameter space and correlations between the model parameters, given the limitations of the simulated models (we use a fiducial model for our SNe~Ia with $x1 = c = 0$, see \cref{sec:simmodels}), we delegate this to a future work.

\begin{figure*}
    \centering
    \includegraphics[width=\textwidth]{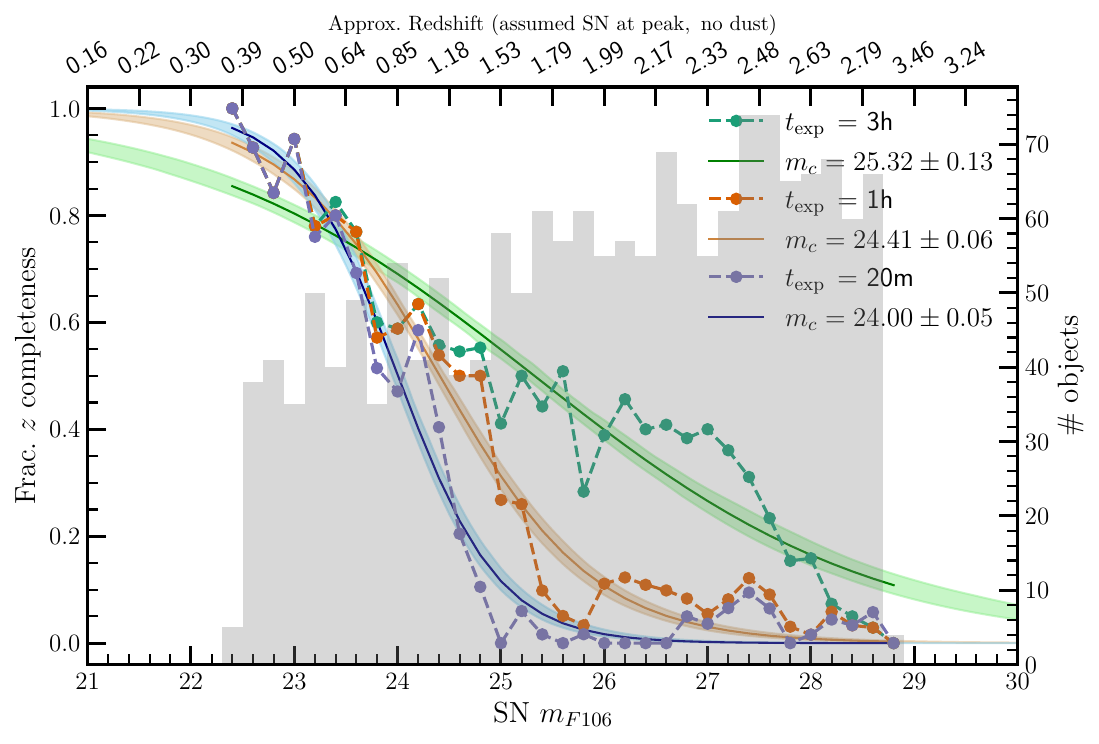}
    \caption{Efficiency of recovering redshifts as a function of SN magnitude in F106 for the entire simulated prism spectroscopic sample of \totalsn\ SNe~Ia. The left ordinate denotes what we refer to as the redshift completeness --- this is the fraction of objects within a given magnitude bin that achieve a redshift accuracy of $\sigma_z = \left|\Delta z/(1+z)\right| \leq 0.01$. The top abscissa provides a guide to the reader as to the approximate redshift for the SN F106 magnitude; the SN redshift shown is derived under the assumption of peak phase and no intervening dust. The background histogram in gray shows the magnitude distribution of the simulated SNe~Ia with the number of objects indicated on the right-hand ordinate. The points joined by dashed lines show the measured redshift completeness from our simulation (compiled from redshifts measured for individual SNe), with colors corresponding to different exposure times as shown in the legend. The solid lines are sigmoid function fits to the measured redshift completeness points. The 50\% redshift completeness magnitude for a given exposure time is denoted by $m_c$ in the legend.}
    \label{fig:completeness}
\end{figure*}

\cref{fig:completeness} displays the efficiency of recovering redshifts as a function of \sn magnitude for our entire simulated sample of \totalsn\ SNe~Ia sources. Within each magnitude bin the figure shows the fraction of spectra that achieve a redshift accuracy of $\leq 0.01$, measured through the absolute value of the redshift residuals $|\sigma_z|$. These redshifts are recovered solely from P127 spectroscopy of the \sn without any reliance on emission or absorption features in the host galaxy spectrum.
While there is an implicit assumption here that the SN~Ia classification efficiency from P127 spectra is 100\% at all redshifts, this is a simplifying assumption we make given that any GO program on \rst will likely have classified a given transient as an \sn from photometric observations, prior to requesting prism spectroscopic observations. From preliminary work (done for HLTDS observations) we know that \sn classification efficiency is $\gtrsim 90$\% for SNe~Ia within $0.3 \lesssim z \lesssim 1.75$, $\sim 65$\% within $1.75 \lesssim z \lesssim 2.0$, and $\sim 20$\% for $z>2$ (Steven Rodney, private communication, using simulated P127 spectra of SNe~Ia which are fit with SNID, \citealt{Blondin2007}). Therefore, while our completeness numbers here should be considered as upper limits, the ability to classify a P127 spectrum as belonging to an \sn only really impacts observations of SNe~Ia at $z>2$ or SNe encountering extreme dust extinction.
 
The longest exposure time of 3\,hr in our simulated observations achieves 50\% efficiency for sources as faint as $\sim$25.5, and has much higher efficiency at  brighter SN magnitudes and lower redshifts. Our shorter exposure times of 20\,min and 1\,hr achieve $>50$\% completeness for SNe~Ia brighter than 24 and 24.4\,mag AB, respectively. Particularly for $z<0.5$ SNe~Ia, this implies that the \rst P127 is quick and efficient at yielding high-S/N spectra and accurate redshifts. Note the gradual flattening of the curves as the exposure time is increased owing to increased numbers of fainter SN spectra being fit successfully, while of course the spectra of the brightest SNe are always fit successfully. For the longest exposure time considered, the gradual transition from high to low completeness also means that the 50\% completeness cutoff (denoted by $m_c$ in the figure legend) is not sharp and the completeness drops to $<25$\% beyond 27.5\,mag AB.

We expect a 3\,hr exposure time to reliably achieve 50\% redshift completeness for all \sn redshifts in the range $1 \leq z \leq 2$. These SNe~Ia are expected to contribute most within cosmological analyses toward improving the dark energy FoM \citep{Albrecht2006}.
Cosmological analyses usually assume that redshift errors have negligible effect on the uncertainty budget in their analysis \citep[e.g.,][]{Betoule2014, Scolnic2018}. We show in this work that this assumption should remain valid with cosmological analyses done using SN~Ia redshifts inferred from \rst prism spectra. We find that the median redshift accuracy, median($|\sigma_z|$), is on the order of a few tenths of a percent for any exposure time that provides S/N $> 5$ (\cref{fig:snr}). How the uncertainty in redshift propagates to the dark energy FoM is part of a separate investigation (Macias et al., in prep.).

SNe~Ia for which a host galaxy cannot be identified \citep[e.g.,][]{Gupta2016}, ``hostless'' SNe, are susceptible to being dropped from cosmological analyses which employ only host galaxy observations to estimate redshifts (e.g., Dark Energy Survey, \citealt{Abbott2019}; Pan-STARRS, \citealt{Jones2018}). Within the framework presented in this paper, i.e., SNe~Ia without host galaxy blended light, we can recover accurate redshifts for \rst prism observations of hostless SNe (right column of \cref{fig:recovery}); therefore, allowing SNe~Ia to be included in cosmological analyses which would otherwise be ignored.

While the primary goal of this work is not to optimize the ratio of P127 to imaging time for the HLTDS, we show a promising outlook for redshift inference from P127 spectra especially for SNe~Ia at $z>1$. \citet{Rose:2021wk} presented several survey strategies with the ratio of prism to imaging time being one of the key unknown variables; the primary reference survey has 25\% prism time amounting to 7.5\,hr per visit split between a wide and a deep tier (900\,s and 3600\,s per object from the wide and deep tiers, respectively). Our 20\,min and 1\,hr exposure times are similar to the exposure time of a single HLTDS pointing, and achieve $> 50$\% completeness for SNe~Ia brighter than 24 and 24.4\,mag AB, respectively. However, if spectra from multiple visits are coadded, then the completeness will improve further. \cref{fig:recovery} shows that the main problem hindering higher levels of redshift completeness is the blending of host galaxy and SN light causing confusion for fitting routines.
In general for any rolling survey, disentangling the SN spectrum from the host galaxy spectrum will be somewhat easier given that the host galaxies will be repeatedly observed, allowing a subtraction of an uncontaminated host spectrum from a host+SN spectrum, as long as data are taken with a wide enough baseline such that SN light has faded.

\subsection{Caveats}
In this section, we discuss three caveats for the work presented here. Firstly, while the simulated spectra in the 2D dispersed images have the appropriate variable dispersion vs.\ wavelength for the \rst prism, the extracted 1D spectra do not yet have a variable dispersion. The 1D spectra have a constant wavelength sampling of 60\,\AA\ per pixel, corresponding to the approximately average dispersion for the \rst prism -- allowing for the variable dispersion in the 1D spectra requires significant restructuring of the \texttt{pyLINEAR} code base and is a work in progress (note the process for 2D simulation and 1D extraction described in \cref{sec:2dsim} and \cref{sec:x1d}). As preliminary confirmation, we note that \cref{fig:recovery} shows that the constant wavelength sampling for the 1D spectra does not hinder our ability to infer accurate redshifts, due to the broad absorption features in \sn spectra.

We also conduct an additional test to ensure that our constant sampling does not significantly affect our recovery of redshift. This is done by resampling the extracted 1D spectra to the expected dispersion of the P127 prism. The resampled spectra are then refit using the same pipeline as in \cref{sec:fitting} and the newly recovered parameters are compared to the previous set of parameters recovered from the spectra with constant sampling.
For the redshifts recovered with the resampled spectra, we observe improvements to a small subset of the sample -- for objects within $z\lesssim2$ an additional $\sim$1--3\% objects pass the redshift accuracy cut of $\left| \sigma_z \right| \leq 0.01$. This translates to an improvement of $\sim0.4$ mag in completeness for the 20 min and 1 hr exposure times and a $\sim0.2$ mag improvement for the longest exposure time. However, we advise caution when interpreting these seemingly large improvements to the completeness depths since the resampling was done post the extraction of the 1D spectra.
The sampling with the variable dispersion shows no significant changes at the highest redshifts but at lower redshifts shows an improvement in the efficiency of recovering redshifts. Therefore, the constant sampling provides a conservative estimate of \rst efficiency of redshift recovery for SNe~Ia and should be employed for current survey planning purposes, whereas redshift recovery with the variable dispersion sampling deserves a more thorough investigation.

Secondly, because our goal is to evaluate redshift accuracies for SNe~Ia from \rst P127 spectra, we do not include other classes of transients in our simulation (e.g.,  other Ia subtypes or core-collapse SNe), and lastly, we do not yet incorporate complex dust geometries in our simulation.

\section{Summary}
We have reported results from the analysis of \totalsn\ simulated \rst P127 spectra of \sn  with exposure times of 1200, 3600, and 10,800\,s. The \rst P127 prism is exceptionally efficient at delivering high-S/N spectra, and therefore is particularly useful for inferring accurate redshifts based on absorption features in the continuum. We recover limiting magnitudes, i.e., $>$50\% completeness, of 24.0, 24.4, and 25.3, for our three exposure times (although see preceding discussion about the steepness of completeness curves) under the requirement of $\sigma_z = (\left|z - z_\mathrm{true} \right|)/(1+z) \leq 0.01$.
The procedure to employ \texttt{pyLINEAR} for simulating and extracting spectra, particularly for the case of the \rst P127 prism, is presented. The entire set of extracted 1D spectra, and simulated and recovered SN properties in tabular form, are available upon request.

\acknowledgements
We are thankful to the anonymous reviewer for their helpful comments and suggestions.
B.A.J.\ is grateful for helpful discussions with members of the \rst Science Investigation Team (SIT) for supernova cosmology, in particular with Steven Rodney, Ryan Foley, and Kevin Wang. A.V.F. received financial support from the Christopher R. Redlich Fund and many individual donors. R.H. acknowledges work supported by NASA under award number 80GSFC21M0002.

The following software packages were used in the analysis presented in this work: \texttt{numpy} \citep{numpy}, \texttt{matplotlib} \citep{matplotlib}, \texttt{scipy} \citep{scipy}, \texttt{astropy} \citep{astropy}, \texttt{emcee} \citep{Foreman-Mackey2013, Foreman-Mackey2019}, \texttt{corner} \citep{Foreman-Mackey2016}, \texttt{pyLINEAR} \citep{Ryan2018}, \texttt{SExtractor} \citep{Bertin1996}, \texttt{SAOImage DS9} \citep{ds9}. This research has made use of NASA’s Astrophysics Data System.

\bibliographystyle{aasjournal}
\bibliography{references}

\appendix
\section{Examples of Parameter Recovery}
\begin{figure*}[t!]
\centering
\includegraphics[width=0.9\textwidth]{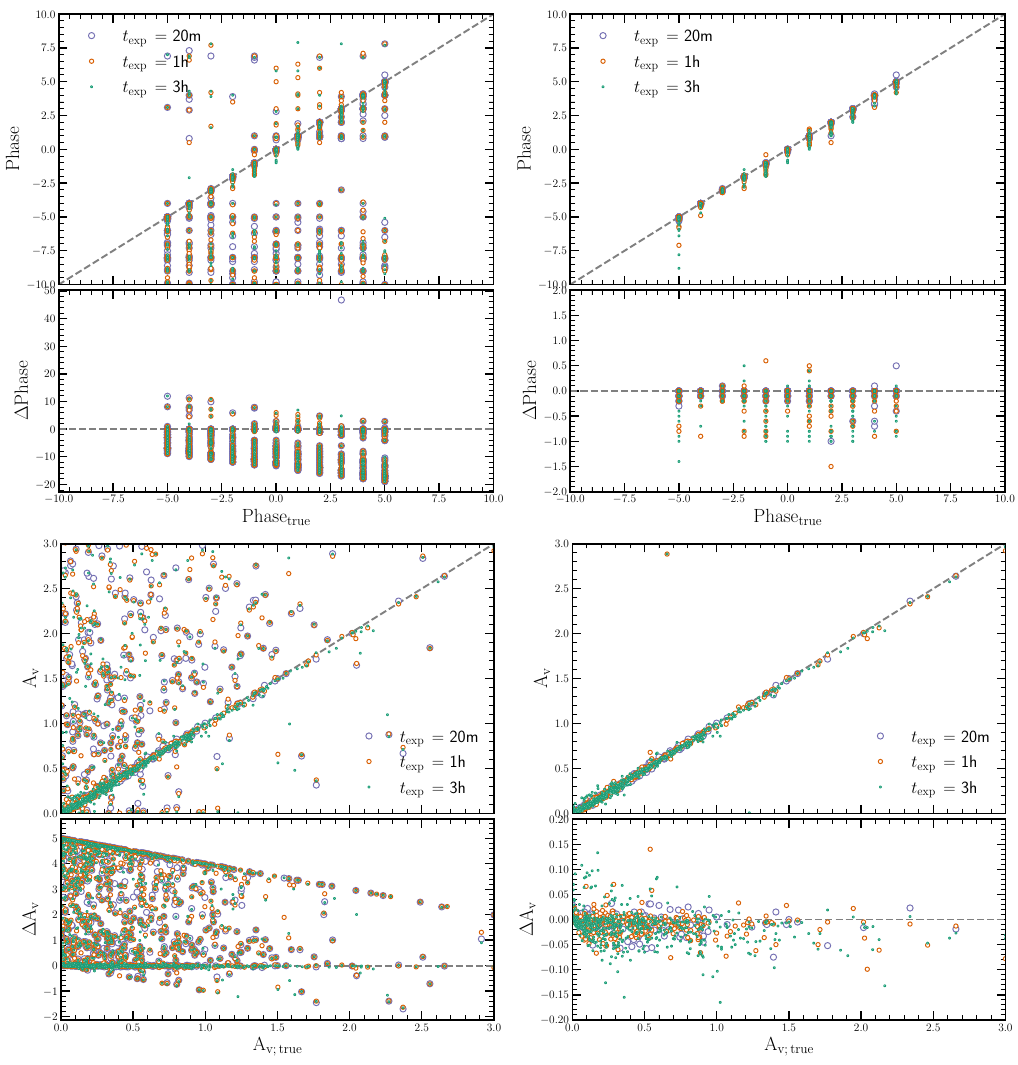}
\caption{The recovery from our fitting pipeline for phase (top row) and dust extinction (bottom row). Similar to \cref{fig:recovery} the left column shows the recovery results for all the SNe~Ia in our simulation, whereas the right column is restricted to only those SNe whose spectra contain no blended light from the host galaxy. The panels show the inferred value vs. the true value on the top and the difference between the inferred and true values on the bottom.}
\label{fig:recovery_appendix}
\end{figure*}

\begin{figure*}
\includegraphics[width=0.98\textwidth]{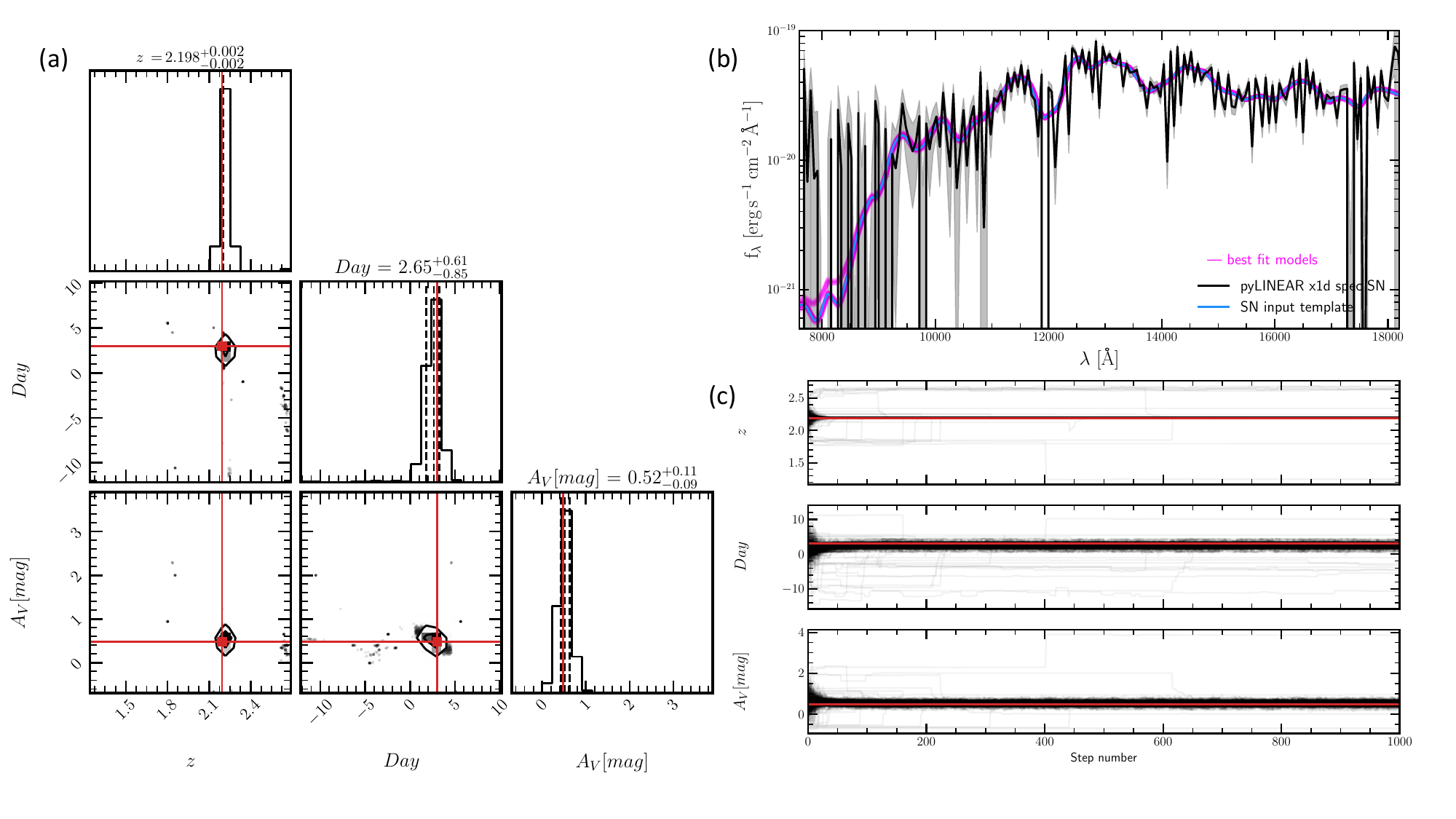}
\caption{An example of a successful fit to a simulated \sn at $z \approx 2.2$ that does not include any blended host galaxy light. Panel (a): Corner plot for the sampled parameter space, where the red lines indicate truth values. The values and uncertainties on top of the panels indicate the estimates from the MCMC sampling. Panel (b): extracted spectrum in black with gray errors and 200 randomly drawn models, in magenta, from the posterior distribution within $\pm 1 \sigma$ of the inferred estimates. The input template for the SN is shown in blue. Panel (c): trace plot for the MCMC walkers, where red lines again indicate truth values. From top to bottom the sub-panels are redshift, phase, and $A_V$. The exposure time is 1\,hr for the extracted 1D spectrum.}
\label{fig:corner_pass}
\end{figure*}

\begin{figure*}
\includegraphics[width=0.98\textwidth]{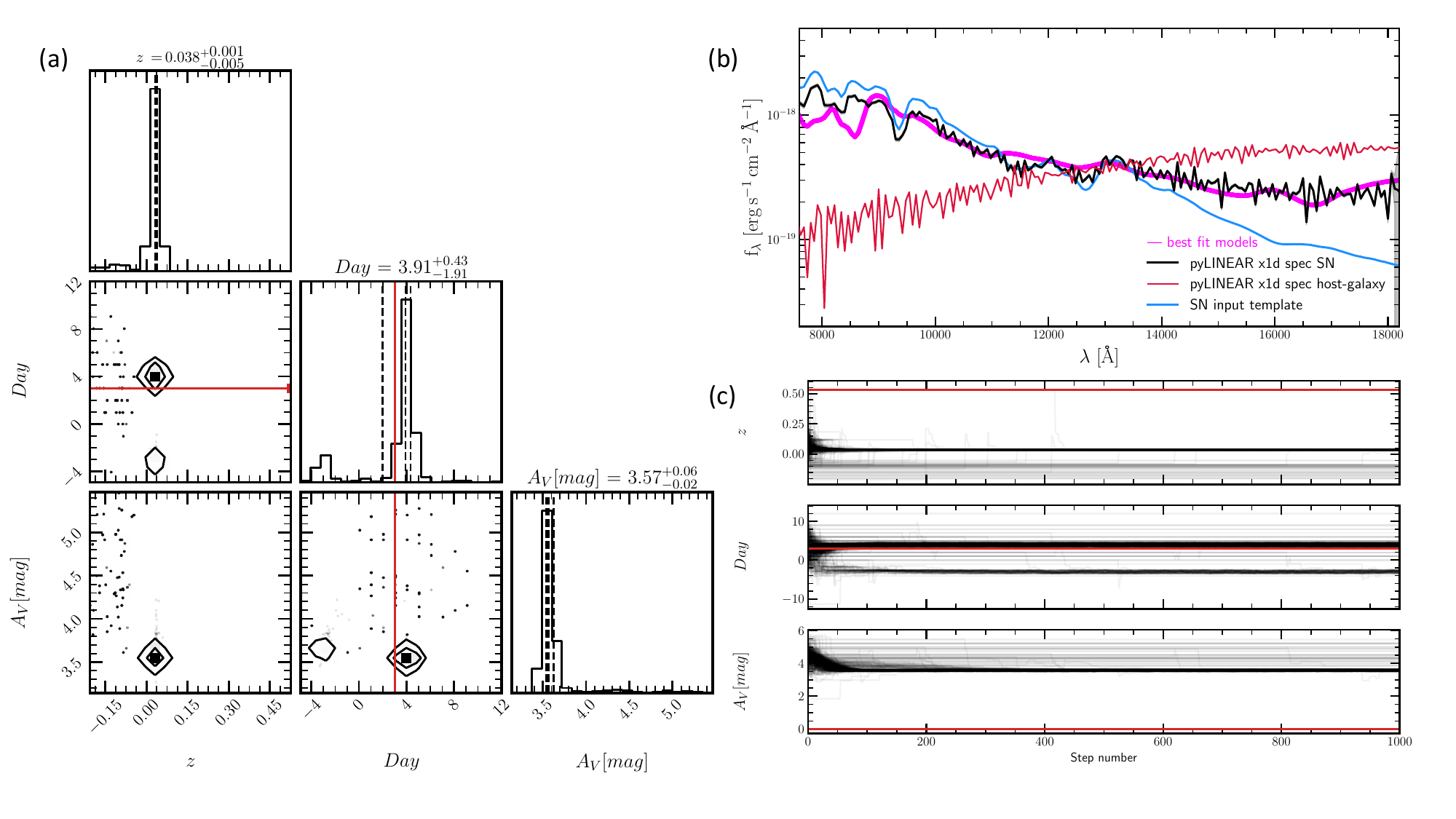}
\caption{Same as \cref{fig:corner_pass} but for an unsuccessful fit due to blended host galaxy light. The extracted 1D spectrum for the host galaxy is also shown (red line) as a reference for the spectrum that is blended with the SN input template. The true values are $z_\mathrm{true} = 0.5288$, $\mathrm{phase_{true}}=3$\,days, and $A^\mathrm{true}_V = 0.02$\,mag. The inferred values are $z = 0.038$, $\mathrm{phase}=3.91$\,days, and $A_V = 3.57$\,mag (also indicated on the corner plot).}
\label{fig:corner_fail}
\end{figure*}

\cref{fig:recovery_appendix} shows the recovery of phase and dust attenuation by our fitting pipeline. 
We also show two representative examples of parameter recovery by our fitting pipeline. Example 1 in \cref{fig:corner_pass} shows the results of a successful fit, whereas example 2 in \cref{fig:corner_fail} shows a common failure mode due to misidentification of absorption features within SNe~Ia spectra that are caused by the red end continuum slope of the host galaxy (\cref{sec:fitting}). Similar to the examples of blending shown in \cref{fig:host_blending} the fitting fails for the extracted spectrum in \cref{fig:corner_fail} due to the $\chi^2$ being dominated by the addition of the host galaxy continuum slope, even though it is clear that the extracted spectrum has the SN absorption features at the correct wavelengths as seen in the input template.

\end{document}